\documentclass[english,aps,english,aip,jcp,dvipsnames,preprint]{revtex4-1}
\usepackage[T1]{fontenc}
\usepackage[latin9]{inputenc}
\usepackage{float}
\usepackage{booktabs}
\usepackage{amsmath}
\usepackage{amssymb}
\usepackage{graphicx}
\usepackage{esint}
\usepackage{subscript}

\makeatletter

\providecommand{\tabularnewline}{\\}

 \@ifundefined{textcolor}{}
 {%
   \definecolor{BLACK}{gray}{0}
   \definecolor{WHITE}{gray}{1}
   \definecolor{RED}{rgb}{1,0,0}
   \definecolor{GREEN}{rgb}{0,1,0}
   \definecolor{BLUE}{rgb}{0,0,1}
   \definecolor{CYAN}{cmyk}{1,0,0,0}
   \definecolor{MAGENTA}{cmyk}{0,1,0,0}
   \definecolor{YELLOW}{cmyk}{0,0,1,0}
 }

\usepackage{braket}

\renewcommand{\textemdash}{---}

\makeatother

\usepackage{babel}
\begin{document}

\title{An Effective Semiclassical Approach to IR Spectroscopy}

\author{Marco \surname{Micciarelli}}
\email{marco.micciarelli@unimi.it}

\author{Fabio \surname{Gabas}}

\author{Riccardo \surname{Conte}}

\author{Michele \surname{Ceotto}}
\email{michele.ceotto@unimi.it}

\affiliation{Dipartimento di Chimica, Università degli Studi di Milano, via C.
Golgi 19, 20133 Milano, Italy.}
\begin{abstract}
We present a novel approach to calculate molecular IR spectra based
on semiclassical molecular dynamics. The main advance from a previous
semiclassical method {[}M. Micciarelli, R. Conte, J. Suarez, M. Ceotto
\emph{J. Chem. Phys.} \textbf{149}, 064115 (2018){]} consists in the
possibility to avoid state-to-state calculations making applications
to systems characterized by sizable densities of vibrational states
feasible. Furthermore, this new method accounts not only for positions
and intensities of the several absorption bands which make up the
IR spectrum, but also for their shapes. We show that accurate semiclassical
IR spectra including quantum effects and anharmonicities for both
frequencies and intensities can be obtained starting from semiclassical
power spectra. The approach is first tested against the water molecule,
and then applied to the 10-atom glycine aminoacid.
\end{abstract}
\maketitle

\section{Introduction}

Infrared (IR) spectroscopy is commonly employed for chemical characterization
owing to some peculiar features which include its cheapness, portability
to different environments, and generally short execution times. The
main features of an IR spectrum, from which important information
about molecular structure and interactions can be obtained, include
the frequencies of vibrational transitions, the shapes of absorption
bands, and their intensities. However, experimental measures can be
of difficult interpretation. This is especially true when the IR spectrum
consists of many overlapping absorption bands, which make it hard
to assign fingerprint and stretch vibrations. It is in this setting
that theoretical simulations of IR spectra may be crucial, allowing
to decompose the spectrum into specific molecular motions. This is
usually achieved first relying on the normal modes of vibration as
a basic harmonic approximation, and then refining the theory by including
anharmonicity effects.\citep{marx_mathias_IRfluxional_2011,gaigeot_gaigeot_floppypeptides_2010}

The frequencies of vibrational transitions can be calculated reliably
and accurately through Semiclassical (SC) molecular dynamics.\citep{Heller_SCspectroscopy_1981,Kaledin_Miller_Timeaveraging_2003,Kaledin_Miller_TAmolecules_2003}
In fact, a SC propagator is able to regain quantum effects from a
classical Hamiltonian dynamics and several SC methods have been introduced
to estimate quantum frequencies of vibration upon calculation of power
spectra.\citep{Heller_FrozenGaussian_1981,Herman_Kluk_SCnonspreading_1984,Miller_Addingquantumtoclassical_2001,Miller_PNAScomplexsystems_2005,Zhang_Pollak_Deeptunneling_2004,Zhang_Pollak_Hybridprefactor_2005,Zhang_Pollak_Noprefactor_2004,Shalashilin_Child_CCS_2004,Shalashilin_Child_Coherentstates_2001}
In practice, a quantum reference state is chosen and the vibrational
eigenenergies are obtained from the Fourier transform of its survival
amplitude. It is then trivial to calculate the vibrational transition
frequencies by difference with respect to the zero-point energy (ZPE).\citep{Ma_Ceotto_SN2reactions_2018,Tamascelli_Ceotto_GPU_2014,DiLiberto_Ceotto_Prefactors_2016}\\
Recent advances have permitted to get SC power spectra of systems
characterized by many degrees of freedom. This has been achieved in
our group by developing innovative methodologies like the multiple
coherent (MC) and the divide-and-conquer (DC) semiclassical initial
value representation (SCIVR). MC SCIVR is based on a tailored choice
of reference state and dynamics initial conditions. In this way, accurate
results are collected running just a few or even a single classical
trajectory.\citep{Ceotto_AspuruGuzik_Multiplecoherent_2009,Ceotto_AspuruGuzik_Curseofdimensionality_2011,Conte_Ceotto_NH3_2013,Ceotto_Tantardini_Copper100_2010}
Such a reduction in the needed computational effort has opened up
the possibility to employ ab initio on-the-fly dynamics and to apply
the semiclassical formalism to systems with many degrees of freedom.\citep{Ceotto_AspuruGuzik_PCCPFirstprinciples_2009,Ceotto_AspuruGuzik_Firstprinciples_2011,Ceotto_Hase_AcceleratedSC_2013,Zhuang_Ceotto_Hessianapprox_2012}
When dealing with high dimensional systems, though, it is not always
possible to get a sensible spectroscopic signal with a full dimensional
SC approach. In the case of a system-bath model, one can employ a
mixed semiclassical approach with an accurate semiclassical propagator
for the system and a less accurate one for the bath.\citep{Buchholz_Ceotto_MixedSC_2016,Buchholz_Ceotto_applicationMixed_2017,Ceotto_Buchholz_SAM_2018}
However, in general, DC SCIVR has been introduced to overcome the
curse of dimensionality issue.\citep{ceotto_conte_DCSCIVR_2017} The
technique, still based on full dimensional classical molecular dynamics,
allows to compute semiclassical power spectra in reduced dimensionality
within a set of appropriately chosen subspaces. The total spectrum
is eventually obtained by collecting the low dimensional spectra calculated
in the subspaces. Some representative applications of these techniques
include fullerene,\citep{ceotto_conte_DCSCIVR_2017} glycine,\citep{Gabas_Ceotto_Glycine_2017}
benzene,\citep{DiLiberto_Ceotto_Jacobiano_2018} water clusters,\citep{Ceotto_watercluster_18}
the protonated glycine dimer and H\textsubscript{2}-tagged protonated
glycine.\citep{Gabas_Ceotto_SupramolecularGlycines_2018}

Regarding band shapes, they arise naturally in dynamical approaches,
like SC ones, from the Fourier transform and include the effect of
any interactions experienced along the dynamics. This is different
from common stick spectra in which a single central transition is
representative of the whole absorption band whose shape is modeled
by means of an \emph{ad hoc} Lorentzian function.

Finally, the accurate estimate of spectral intensities is still a
partially open issue and the missing tile for a complete semiclassical
simulation of IR spectra. The straightforward approach to the problem
deals with the calculation of the dipole autocorrelation, since its
Fourier transform returns the correct estimates for both transition
frequencies and intensities.\citep{Heller_SCspectroscopy_1981,Wehrle_Vanicek_Oligothiophenes_2014,Patoz_Vanicek_Photoabs_Benzene_2018,Wehrle_Vanicek_NH3_2015,Zambrano_Vanicek_Cellulardephasing_2013,Sulc_Vanicek_CellularDephasing_2012,Jiri_Vanicek_ultrafastReview_18,Jiri_Vanicek_phenylradical_18,Jiri_Vanicek_Chem_Phys_2018,Tatchen_Pollak_Onthefly_2009}
Strictly speaking, though, the presence of two propagators leads semiclassically
to a double phase-space integration, which makes the calculation very
hard to converge due to the oscillations produced by the phase differences
between trajectories with different initial conditions.\\ Several
methods have been developed to try to overcome this issue. A possible
strategy consists in easing the calculation by means of a filter able
to damp the oscillations, hopefully without spoiling the results.\citep{Filinov_Filter_1986,Makri_Miller_Filinov_1988,Church_Ananth_Filinov_2017}
Another possibility is represented by the linearization approximation,
which leads to a semiclassical expression based on a single phase
space integration and formally equal to its classical counterpart,
but with Wigner functions replacing classical ones.\citep{Sun1997,Sun1998,Sun1998-1,Liu_Miller_linearizedSCIVR_2007,Liu_Miller_VariouslinearizedSCIVR_2008}
This kind of approximation has the drawback that it is unable to account
for quantum coherence between distinct trajectories and is affected
by the zero point energy leakage problem, which is not the case for
the original SCIVR formulation.\citep{Buchholz_Ceotto_leakage_2018}
A more accurate approach would be based on the rearrangement of the
dipole autocorrelation in a suitable way for application of the forward-backward
SC formula obtained by stationary phase approximating the primitive
SC expression. In this way a single phase space integration is needed,
quantum coherences are included, and oscillations are quenched due
to the evolution of trajectories first forward and then backward in
time.\citep{Makri_Thompson_FBIVR_1998,shao_makri_backfrowardnopref_1999,Sun_Miller_FBIVR_1999,MFierro_Pollak_FBIVR_2006,Pollak_MFierro_FBIVR_2007}
In all cases, though, the challenge is to go beyond model systems
and to be able to get semiclassical IR spectra even for molecules
of sizable dimensionality, i.e. the goal already reached for power
spectra.

For this purpose a SC methodology able to reproduce IR spectra has
been introduced very recently.\citep{Micciarelli_Ceotto_SCwavefunctions_2018}
It is based on the calculation of oscillator strengths from vibrational
eigenfunctions represented as linear combinations of harmonic functions.
The coefficients of the linear combinations are obtained after collecting
information from an appropriate set of SC power spectrum simulations
in which harmonic states are employed as reference states. The approach
takes advantage from the techniques developed for power spectra, and
applications to systems with many degrees of freedom are in principle
feasible. Results for the water molecule are indeed in excellent agreement
with the quantum benchmark, but the method has the drawback to require
calculation of contributions from all states involved in the transitions
that make up the several absorption bands in the IR spectrum, which
is unpractical for systems characterized by large densities of vibrational
states.

The principal aim of the present paper is to take SC IR spectroscopy
to the same level of applicability as SC power spectrum investigations.
The goal is achieved by decomposing the IR spectrum into a sum of
dynamical correlation functions, which can be calculated from SC power
spectra and allows to avoid the undesired state-to-state computations.
The paper outline is as follows: In Section \ref{sec:Theory} we detail
the theory behind the new method; Section \ref{sec:RESULTS-AND-DISCUSSION}
is devoted to two representative applications. Specifically, application
to water yields results of excellent accuracy, while application to
the high energy fundamentals of glycine points out the importance
of electrical and mechanical anharmonicity, and demonstrates the possibility
to get reliable IR spectra for higher dimensional systems. A summary
and some conclusions end the paper.

\section{Theory\label{sec:Theory}}

\textbf{Eigenvalues, eigenfunctions and transition dipoles }For a
molecular system governed by the vibrational Hamiltonian operator

\begin{align}
\hat{H}=\hat{T}+\hat{V}\label{eq:BO_Hamiltonian}
\end{align}
with potential energy surface (PES) $V(\boldsymbol{R})$ given by
the electronic Born-Oppenheimer adiabatic energy, we are interested
in studying the spectral decomposition of $\hat{H}$ in terms of vibrational
bound states beyond the harmonic approximation, i.e. in solving the
eigenvalue problem

\begin{align}
\hat{H}\ket{e_{n}}=E_{n}\ket{e_{n}}.\label{eq:H-spectrum}
\end{align}
An analytical solution to Eq.(\ref{eq:H-spectrum}) is in general
not available, so numerical and approximate strategies are needed.
For this purpose, the theoretical foundation of the formalism employed
in this work lies on the adoption of the time propagation operator

\begin{align}
\mathcal{\hat{{P}}}(t)=e^{-\frac{i}{\hbar}\hat{H}t}=\sum_{n}e^{-\frac{i}{\hbar}E_{n}t}\ket{e_{n}}\bra{e_{n}}\label{eq:propagator-exact}
\end{align}
to compute the time dependent survival amplitude of an arbitrary reference
state $\ket{\chi}$

\begin{align}
I_{\chi}(t) & \equiv\bra{\chi}\hat{\mathcal{P}}(t)\ket{\chi}=\nonumber \\
 & =\sum_{n}e^{-\frac{i}{\hbar}E_{n}t}~\left|\braket{\chi|e_{n}}\right|^{2},\label{eq:Rec-T-overlap}
\end{align}
where the second equality is obtained upon introduction of the representation
of the propagator in the basis of the Hamiltonian eigenvectors.

Eq. (\ref{eq:Rec-T-overlap}) implies that both squared projections
$\left|\braket{\chi|e_{n}}\right|^{2}$ of the reference state onto
the eigenvectors and eigenvalues $E_{n}$ can be determined respectively
from peak amplitudes and positions of the following power spectrum

\begin{flalign}
\tilde{I}_{\chi}(E)\! & =\frac{1}{2\pi\hbar}\int_{-\tau}^{\tau}\!\!dtI_{\chi}(t)e^{\frac{i}{\hbar}Et}=\nonumber \\
 & =\!\frac{1}{\pi\hbar}~Re\left[\int_{0}^{\tau}dt~I_{\chi}(t)e^{\frac{i}{\hbar}Et}\right]=\nonumber \\
 & =\sum_{n}\left|\braket{\chi|e_{n}}\right|^{2}\mathcal{D}(E-E_{n};\Gamma_{\tau}).\label{eq:Rec-T-ovelap-power}
\end{flalign}
The dynamical convolution function $\mathcal{D}$ is a delta-nascent
function centered on $E_{n}$ with amplitude $\Gamma_{\tau}$ approaching
zero as $\tau\rightarrow\infty$. It has the analytical form of a
sinc function if the exact propagator is employed.

While eigenvalues are obtained straightforwardly, determination of
eigenvectors requires a more elaborated strategy. In a recent paper\citep{Micciarelli_Ceotto_SCwavefunctions_2018}
we have shown how vibrational eigenstates can be derived from SC power
spectra. For a generic system with $N_{v}$ vibrational degrees of
freedom, the starting point is the representation of the eigenvectors
as a linear combination of $N_{v}$-dimensional harmonic states $\ket{\phi_{\mathbf{K}}}$

\begin{align}
\ket{e_{n}}=\sum_{\textbf{K}=1}^{N}C_{n,\mathbf{K}}\ket{\phi_{\mathbf{K}}}.\label{eq:en-expansion}
\end{align}
$N$ is the number of states in the basis set, $C_{n,\mathbf{K}}=\left<\phi_{\mathbf{K}}|e_{n}\right>$
are real coefficients, and $\ket{\phi_{\mathbf{K}}}$ is the generic
$\mathbf{K}$-th element of the complete and orthonormal N-dimensional
basis set $\left\{ \ket{\textbf{\ensuremath{\phi}}_{{\mathbf{K}}}}\right\} $.
Each element $\ket{\phi_{\mathbf{K}}}$ of the basis set is obtained
as the Hartree product of one-dimensional harmonic states

\begin{align}
\ket{\textbf{\ensuremath{\phi_{\mathbf{K}}}}} & =\ket{\ensuremath{\phi_{K_{1}}},\phi_{K_{2}},~\dots~,\phi_{K_{Nv}}}=\nonumber \\
 & =\ket{\ensuremath{\phi_{K_{1}}}}~\dots\ \ket{\phi_{K_{Nv}}},\label{eq:harm-basis-ket}
\end{align}
and

\begin{align}
\phi_{K_{\alpha}}(Q_{\alpha}) & =\braket{Q_{\alpha}|\phi_{K_{\alpha}}}=\frac{1}{\sqrt{2^{K_{\alpha}}K_{\alpha}!}}\left(\frac{\omega_{\alpha}}{\pi\hbar}\right)^{\frac{1}{4}}\times\nonumber \\
 & e^{-\frac{\omega_{\alpha}Q_{\alpha}^{2}}{2\hbar}}h_{K_{\alpha}}\left(\sqrt{\frac{\omega_{\alpha}}{\hbar}}~Q_{\alpha}\right),\label{eq:harm-basis-x}
\end{align}
where $Q_{\alpha}=q_{\alpha}-q_{eq,\alpha}$, $\omega_{\alpha}$,
and $K_{\alpha}$ are, respectively, the $\alpha$-th normal mode
coordinate displacement from equilibrium, frequency, and quantum number.
$h_{K_{\alpha}}$ indicates the $K_{\alpha}$-th order Hermite polynomial
in the variable $Q_{\alpha}$. The square modulus of coefficients
$C_{n,\textbf{K}}$ can be computed considering that they are proportional
to the intensity, at the eigenvalue of the vibrational Hamiltonian,
of the Fourier transform of the survival amplitude with reference
harmonic state $\phi_{{\bf K}}$, i.e. $|C_{n,\mathbf{K}}|^{2}\propto\tilde{I}_{\phi_{\mathbf{K}}}(E_{n})$.
As shown in details in our previous work,\citep{Micciarelli_Ceotto_SCwavefunctions_2018}
the signed coefficients in Eq. (\ref{eq:en-expansion}) can be calculated
from survival amplitudes using the following working formula

\begin{align}
C_{n,\mathbf{K}}=\frac{\Delta\tilde{I}_{\phi_{\textbf{0}},\phi_{\textbf{K}}}(E_{n})}{2\sqrt{\tilde{I}_{\phi_{{\boldsymbol{0}}}}(E_{n})}},\label{eq:TA-coeff-last}
\end{align}
where $\phi_{\mathbf{0}}$ is the harmonic ground state, $\tilde{I}_{\phi_{{\boldsymbol{k}}}}(E_{n})$
is the value at energy $E_{n}$ of the power spectrum obtained with
the harmonic state $\ket{\phi_{\mathbf{K}}}$, and

\begin{align}
\Delta\tilde{I}_{\phi_{\mathbf{K}_{1}},\phi_{\mathbf{K}_{2}}}(E)\equiv & \tilde{I}_{\phi_{\mathbf{K}_{1}}+\phi_{\mathbf{K}_{2}}}(E)-\tilde{I}_{\phi_{\mathbf{K}_{1}}}(E)-\tilde{I}_{\phi_{\mathbf{K}_{2}}}(E).\label{eq:survival-harmo-Delta}
\end{align}

The same coefficients can be exploited to evaluate the transition
dipole between an initial state $\ket{e_{n}}$ and a final state $\ket{e_{m}}$

\begin{align}
\braket{e_{n}|\hat{\mu}_{0N}^{\epsilon}|e_{m}}=\sum_{\mathbf{K},\mathbf{K}'}C_{n,\mathbf{K}}C_{m,\mathbf{K}'}\bra{\mathbf{\phi_{\mathbf{K}}}}\hat{\mu}_{0N}^{\epsilon}\ket{\phi_{\mathbf{K}^{\prime}}},\label{eq:dip-on-h-basis}
\end{align}
which is needed for calculating the absorption spectrum. In Eq.(\ref{eq:dip-on-h-basis})
$\epsilon=x,y,z$ and $\hat{\pmb{\mu}}_{0N}(\mathbf{R})=\hat{\pmb{\mu}}_{N}(\mathbf{R})+\hat{\pmb{\mu}}_{e0}(\mathbf{R})$
is the molecular dipole made of two contributions: $\pmb{\mu}_{N}({\bf R})=\sum_{i}Z_{i}\mathbf{R}_{i}$
is the nuclear part; $\pmb{\mu}_{e0}(\pmb{R})~=~\int d\mathbf{r}~|\varphi_{0}(\mathbf{r};\mathbf{R})|^{2}\pmb{\mu}_{e}(\mathbf{r})$
is the electronic dipole with $\varphi_{0}(\mathbf{r};\mathbf{R})$
representing the adiabatic electronic ground state wavefunction for
a given nuclear configuration. Calculation of the nuclear contribution
is trivial, while the electronic one requires a Monte Carlo estimate.\citep{Micciarelli_Ceotto_SCwavefunctions_2018}
The latter can be avoided by means of the widely employed linearization
approximation to the dipole

\begin{align}
\pmb{\mu}_{0N}(\mathbf{q}) & -\pmb{\mu}_{0N}(\mathbf{q}_{eq})\simeq\sum_{\alpha=1}^{N_{v}}\left.\frac{\partial\pmb{\mu}_{0N}}{\partial q_{\alpha}}\right|_{\mathbf{q}_{eq}}(q_{\alpha}-q_{eq,\alpha}).\label{eq:DH-approx}
\end{align}
Eq.(\ref{eq:dip-on-h-basis}) then can be rearranged as

\begin{align}
\braket{e_{n}|\hat{\mu}_{0N}^{\epsilon}|e_{m}}=\sum_{\mathbf{K},\mathbf{K}'}C_{n,\mathbf{K}}C_{m,\mathbf{K}'}\sum_{\alpha}Z_{\epsilon,\alpha}\bra{\phi_{K_{\alpha}}}\hat{Q}_{\alpha}\ket{\phi_{K'_{\alpha}}},\label{eq:dip-on-h-basis_2}
\end{align}
where $Z_{\epsilon,\alpha}=\left.\frac{\partial\mu_{0N}^{\epsilon}}{\partial q_{\alpha}}\right|_{\mathbf{q}_{eq}}$
is a quantity easy to compute and routinely returned by the most popular
electronic structure softwares. Eq.(\ref{eq:dip-on-h-basis_2}) represents
the fully anharmonic estimate of transition dipoles within the linearized
dipole approximation. Other, more approximate formulae can be adopted
to ease the calculation. For instance, the initial state $\ket{e_{n}}$
can be approximated by a single harmonic state $\ket{\phi_{\mathbf{\bar{K}}}}$,
i.e. $\ket{e_{n}}\simeq\sum_{\mathbf{K}}\delta_{\mathbf{K}\mathbf{\bar{K}}}\ket{\phi_{\mathbf{K}}}$.
This leads to

\begin{equation}
\braket{e_{n}|\hat{\mu}_{0N}^{\epsilon}|e_{m}}=\sum_{\mathbf{K}'}C_{m,\mathbf{K}'}\sum_{\alpha}Z_{\epsilon,\alpha}\bra{\phi_{\bar{K}_{\alpha}}}\hat{Q}_{\alpha}\ket{\phi_{K'_{\alpha}}}.\label{eq:semianharmonic_trans_dipole}
\end{equation}
We refer to Eq.(\ref{eq:semianharmonic_trans_dipole}) as the semi-anharmonic
transition dipole. Finally, the harmonic approximation can be invoked
also for the final state, so that $\ket{e_{m}}\simeq\ket{\phi_{\mathbf{\bar{{\bf K^{'}}}}}}$
and the transition dipole takes the simpler form

\begin{equation}
\braket{e_{n}|\hat{\mu}_{0N}^{\epsilon}|e_{m}}=\sum_{\alpha}Z_{\epsilon,\alpha}\bra{\phi_{\bar{K}_{\alpha}}}\hat{Q}_{\alpha}\ket{\phi_{\bar{K'}_{\alpha}}}.\label{eq:harmonic_transition_dipole}
\end{equation}
This is the case commonly known as the harmonic electrical approximation,
which is often coupled in basic spectroscopy calculations to its mechanical
counterpart (i.e. the harmonic estimate of frequencies) under the
collective name of double harmonic approximation.

A matrix element like the one in Eq.(\ref{eq:harmonic_transition_dipole})
can be readily evaluated. In fact

\begin{align}
 & \bra{\mathbf{\phi_{K}}}\hat{Q}_{\alpha}\ket{\phi_{\mathbf{K}'}}=\left(\ \prod_{\beta\ne\alpha}^{N_{v}}\delta_{K_{\beta},K'_{\beta}}\right)\sqrt{\frac{1}{2\omega_{\alpha}}}\times\nonumber \\
 & \left(\delta_{K_{\alpha},K'_{\alpha}+1}\sqrt{K_{\alpha}^{'}+1}+\delta_{K_{\alpha},K'_{\alpha}-1}\sqrt{K_{\alpha}^{'}}\right),\label{eq:selection-rules}
\end{align}

\noindent which is obtained starting from $\hat{Q}_{\alpha}=\sqrt{\frac{1}{2\omega_{\alpha}}}\left(\hat{a}_{\alpha}^{\dagger}+\hat{a}_{\alpha}\right)$,
with $\hat{a}_{\alpha}^{\dagger}$ and $\hat{a}_{\alpha}$ being the
harmonic oscillator creation and annihilation operators, respectively,
for normal mode $\alpha$. Eq.(\ref{eq:selection-rules}) incorporates
the harmonic selection rules, which permit to simplify Eq.(\ref{eq:dip-on-h-basis_2})
by neglecting many zero-valued terms

\begin{align}
\braket{e_{n}|\hat{\mu}_{0N}^{\epsilon}|e_{m}}=\mu_{nm}^{\epsilon}=\sum_{\mathbf{K}}C_{n,\mathbf{K}}\sum_{\alpha}C_{m,\mathbf{K}^{(\alpha)}}\mu_{\epsilon,\mathbf{K}^{(\alpha)}.}\label{eq:dip-with-sel-rules}
\end{align}
The second sum in Eq.(\ref{eq:dip-on-h-basis_2}) is restricted in
Eq.(\ref{eq:dip-with-sel-rules}) to the $2N_{v}$ basis set elements
obtained by exciting or de-exciting the $\alpha$-th degree of freedom,
i.e. the states $\ket{\phi_{\mathbf{K}^{(\alpha)}}}$ such that $\ket{\phi_{\mathbf{K}^{(\alpha)}}}\propto\hat{a}_{\alpha}^{\dagger}\ket{\phi_{\mathbf{K}}}$
or $\ket{\phi_{\mathbf{K}^{(\alpha)}}}\propto\hat{a}_{\alpha}\ket{\phi_{\mathbf{K}}}$.
Furthermore

\begin{align}
\mu_{\epsilon,\mathbf{K}^{(\alpha)}}=\begin{cases}
Z_{\epsilon,\alpha}\sqrt{\frac{K_{\alpha}+1}{2\omega_{\alpha}}} & \quad\text{if }\ket{\phi_{\mathbf{K}^{(\alpha)}}}\propto\hat{a}_{\alpha}^{\dagger}\ket{\phi_{\mathbf{K}}}\\
Z_{\epsilon,\alpha}\sqrt{\frac{K_{\alpha}}{2\omega_{\alpha}}} & \quad\text{if }\ket{\phi_{\mathbf{K}^{(\alpha)}}}\propto\hat{a}_{\alpha}\ket{\phi_{\mathbf{K}}}\quad\text{and }\quad\hat{a}_{\alpha}\ket{\phi_{\mathbf{K}}}\neq0\\
0 & \quad\text{if }\ket{\phi_{\mathbf{K}^{(\alpha)}}}\propto\hat{a}_{\alpha}\ket{\phi_{\mathbf{K}}}\quad\text{and }\quad\hat{a}_{\alpha}\ket{\phi_{\mathbf{K}}}=0
\end{cases}
\end{align}
\textbf{The IR spectrum from power spectra }Our target is the calculation
of the following spectrum

\begin{align}
S_{\epsilon,n}(\omega)=\sum_{m\neq n}~|\mu_{nm}^{\epsilon}|^{2}\,\omega\,\mathcal{D}(\omega-E_{m}+E_{n};\Gamma_{\tau}),\label{eq:spec-n-bro}
\end{align}
in which the sum runs over all the spectral lines~from the $n$-th
to the $m$-th vibrational states. This is an approximate expression
for the IR absorption of a system initially lying on the $n$-th pure
Hamiltonian eigenstate and perturbed by a radiation of frequency $\omega$
polarized along the $\epsilon$ direction, which takes the energy
of the system to $E=\omega+E_{n}$. The spectral lines are broadened
over a finite range of energies by means of the dynamical convolution
function $\mathcal{D}$. Details on this approach and derivation of
Eq.(\ref{eq:spec-n-bro}) are presented in Appendix A.

It is convenient to start by considering the following IR spectrum
in which the transition dipoles are treated by means of the semi-anharmonic
approximation (see Eq.(\ref{eq:semianharmonic_trans_dipole}))

\begin{align}
S_{\epsilon,\phi_{\mathbf{\bar{K}}}}(\omega)=\sum_{m}~|\mu_{\mathbf{\bar{K}}m}^{\epsilon}|^{2}\,\omega\,\mathcal{D}(\omega-E_{m}+E_{n};\Gamma_{\tau}),\label{eq:spec-K-bro}
\end{align}
where

\begin{align}
\mu_{\mathbf{\bar{K}}m}^{\epsilon}=\braket{\phi_{\mathbf{\bar{K}}}|~\hat{\mu}_{0N}^{\epsilon}~|~e_{m}}.\label{eq:mu-Km}
\end{align}
Inserting Eq.(\ref{eq:dip-with-sel-rules}) (with $C_{n\boldsymbol{\bar{K}}}=\delta_{n,\boldsymbol{\bar{K}}}$)
into Eq.(\ref{eq:spec-K-bro}) and expanding the square in the sum,
we get

\begin{align}
S_{\epsilon,\phi_{\mathbf{\bar{K}}}}(\omega)= & \sum_{m}\Bigg(\sum_{\alpha}C_{m,\bar{\mathbf{K}}^{(\alpha)}}^{2}(\mu_{\mathbf{\bar{K}}^{(\alpha)}}^{\epsilon})^{2}+\nonumber \\
+ & 2\sum_{\beta<\alpha}C_{m,\bar{\mathbf{K}}^{(\alpha)}}C_{m,\bar{\mathbf{K}}^{(\beta)}}\mu_{\mathbf{\bar{K}}^{(\alpha)}}^{\epsilon}\mu_{\mathbf{\bar{K}}^{(\beta)}}^{\epsilon}\Bigg)\,\omega\,\mathcal{D}(\omega-E_{m}+E_{n};\Gamma_{\tau}),\label{eq:spec-Km-sc-2}
\end{align}
where the indexes $\alpha$ and $\beta$ run over the normal modes.
Proper combination of Eq.(\ref{eq:Rec-T-ovelap-power}), Eq.(\ref{eq:en-expansion}),
and Eq.(\ref{eq:survival-harmo-Delta}) permits to rearrange Eq.(\ref{eq:spec-Km-sc-2})
in terms of power spectra only. In fact, by means of

\begin{align}
\tilde{I}_{\phi_{{\boldsymbol{K}}}}(E) & =\sum_{m}|C_{m,\boldsymbol{K}}|^{2}\mathcal{D}(E-E_{m};\Gamma_{\tau})\nonumber \\
\Delta\tilde{I}_{\phi_{\mathbf{K}_{1}},\phi_{\mathbf{K}_{2}}}(E) & =\sum_{m}2C_{m,\boldsymbol{K}_{1}}C_{m,\boldsymbol{K}_{2}}\mathcal{D}(E-E_{m};\Gamma_{\tau}),\label{eq:surv-harm-states}
\end{align}
and inverting in Eq.(\ref{eq:spec-Km-sc-2}) the sum over $m$ (that
we do not want to compute) and the sum over $\alpha$ (that we want
to keep explicit), after some straightforward algebra we get

\begin{align}
S_{\epsilon,\phi_{\mathbf{\bar{K}}}}(\omega)=\omega\sum_{\alpha}\Bigg[\mu_{\mathbf{\bar{K}}^{(\alpha)}}^{2}\tilde{I}_{\phi_{\mathbf{\bar{K}}^{(\alpha)}}}(\omega+E_{n})+\sum_{\beta<\alpha}\mu_{\mathbf{\bar{K}}^{(\alpha)}}\mu_{\mathbf{\bar{K}}^{(\beta)}}\Delta\tilde{I}_{\phi_{\mathbf{\bar{K}}^{(\alpha)}},\phi_{\mathbf{\bar{K}}^{(\beta)}}}(\omega+E_{n})\Bigg].\label{eq:spec-Km-sc-finale}
\end{align}
It is worth noting that in Eq.(\ref{eq:spec-Km-sc-finale}) the dependence
of the power spectra on the radiation frequency has been explicitly
indicated, as derived from the anticipated key relation $E=\omega+E_{n}$.

Eventually, the (fully anharmonic) IR spectrum defined in Eq.(\ref{eq:spec-n-bro})
can be obtained (see Appendix B for derivation details) as

\begin{align}
S_{\epsilon,n}(\omega)= & \sum_{\mathbf{K}}C_{n,\mathbf{K}}^{2}S_{\epsilon,\phi_{\mathbf{K}}}(\omega)+\nonumber \\
+\,\omega & \sum_{{\mathbf{K'}}<{\mathbf{K}}}\sum_{\alpha,\alpha'}C_{n,\mathbf{K}}C_{n,\mathbf{K'}}\Delta\tilde{I}_{\phi_{\mathbf{K}^{(\alpha)}},\phi_{\mathbf{K'}^{(\alpha')}}}(\omega+E_{n})\mu_{\mathbf{K}^{(\alpha)}}^{\epsilon}\mu_{\mathbf{K'}^{(\alpha')},}^{\epsilon}\label{eq:final-sc-ir-spec}
\end{align}
where in the sum over ${\mathbf{K'}}<{\mathbf{K}}$ the elements of
the basis set are sorted in some arbitrary way. Eq.(\ref{eq:spec-Km-sc-finale})
and Eq.(\ref{eq:final-sc-ir-spec}) demonstrate that a state-to-state
computation is not required, and that, if power spectra are calculated
exactly, it is only the fully anharmonic estimate that demands for
coefficients (i.e. for knowledge of the eigenfunction of the initial
state).

\textbf{Semiclassical power spectra }In semiclassical dynamics the
quantum propagator is usually approximated by means of the Herman-Kluk
(HK) expression

\begin{align}
\hat{\mathcal{P}}(t)\propto\int\int d\mathbf{Q}_{0}d\mathbf{p}_{0}~C_{t}(\mathbf{Q}_{0},\mathbf{p}_{0})e^{\frac{i}{\hbar}S_{t}(\mathbf{Q}_{0},\mathbf{p}_{0})}\ket{\mathbf{Q}_{t},\mathbf{p}_{t}}\bra{\mathbf{Q}_{0},\mathbf{p}_{0}},
\end{align}
where $\mathbf{Q}_{t}$ and $\mathbf{p}_{t}$ are the classical normal
mode displacement and momentum vectors at time t, obtained from the
classical propagation of the trajectory started at $(\mathbf{Q}_{0},\mathbf{p}_{0})$
under the classical vibrational Hamiltonian. $\ket{\mathbf{Q}_{t},\mathbf{p}_{t}}$
are coherent states of the form

\begin{align}
\braket{\mathbf{x}|\mathbf{Q}_{t},\mathbf{p}_{t}}=\left(\frac{det(\boldsymbol{\gamma})}{\pi}\right)^{\frac{N_{v}}{4}}e^{-\frac{1}{2}(\mathbf{x}-\mathbf{Q}_{t})^{T}\boldsymbol{\gamma}(\mathbf{x}-\mathbf{Q}_{t})+\frac{i}{\hbar}\mathbf{p}_{t}^{T}(\mathbf{x}-\mathbf{Q}_{t})},\label{eq:def-coherents}
\end{align}
where $\boldsymbol{\gamma}$ is a $N_{v}\times N_{v}$ diagonal matrix
with diagonal elements equal to the harmonic frequencies $\{\omega_{\lambda}\}_{\lambda=1}^{N_{v}}$;
$S_{t}$ is the classical action at time $t$ computed along the trajectory,
and, finally, $C_{t}(\mathbf{Q}_{0},\mathbf{p}_{0})$ is the HK prefactor
at time $t$ that accounts for second order quantum fluctuations around
each classical path and which is defined as

\begin{equation}
C_{t}\left(\mathbf{Q}_{0},\mathbf{p}_{0}\right)=\sqrt{\frac{1}{2^{N_{v}}}\left|\frac{\partial\mathbf{Q}_{t}}{\partial\mathbf{Q}_{0}}+\boldsymbol{\gamma}^{-1}\frac{\partial\mathbf{p}_{t}}{\partial\mathbf{p}_{0}}\boldsymbol{\gamma}-i\hbar\frac{\partial\mathbf{Q}_{t}}{\partial\mathbf{p}_{0}}\boldsymbol{\gamma}+\frac{i\boldsymbol{\gamma}^{-1}}{\hbar}\frac{\partial\mathbf{p}_{t}}{\partial\mathbf{Q}_{0}}\right|}.\label{eq:prefactor}
\end{equation}
The multi-dimensional integral over the initial phase space conditions
is usually performed by means of Monte Carlo techniques, and the method
has been applied successfully in many instances, yielding accurate
results.\citep{Bonella_Coker_Linearizedpathintegral_2005,Grossmann_SChybrid_2006,Tao_Miller_Tdepsampling_2011,Tao_Miller_Tdepsampling_2012,Conte_Ceotto_book_chapter_2019} However, the number
of different classical trajectories to run is often prohibitively
high for an effective interface to ab initio on-the-fly evaluations
of energies and gradients.

The computational cost required for this kind of simulations can be
much decreased by employing Kaledin and Miller's time average filter
with separable approximation to the prefactor.\citep{Kaledin_Miller_Timeaveraging_2003}
This approximation consists in imposing that the amplitude of the
HK prefactor is constant in time, i.e.

\begin{align}
C_{t}\left(\mathbf{Q}_{0},\mathbf{p}_{0}\right)\simeq e^{i\phi_{t}\left(\mathbf{Q}_{0},\mathbf{p}_{0}\right)},
\end{align}
a condition that is exactly fulfilled in the case of the harmonic
oscillator. In this way, the TA SCIVR power spectrum is

\begin{flalign}
\tilde{I}_{\textbf{\ensuremath{\chi}}}(E)\! & \propto\int\!\!\!\!\int d\mathbf{Q}_{0}d\mathbf{p}_{0}~\dfrac{1}{\tau}\left|\int_{0}^{\tau}dt\braket{\chi|\mathbf{Q}_{t},\mathbf{p}_{t}}e^{i[S_{t}(\mathbf{Q}_{0},\mathbf{p}_{0})+\phi_{t}(\mathbf{Q}_{0},\mathbf{p}_{0})+Et]/\hbar}\right|^{2}\\
\nonumber 
\end{flalign}
Calculation of the time averaged power spectrum still requires to
perform a multidimensional integration but yields converged results
orders of magnitude faster with a loss in accuracy in peak positions
of just a few $\text{cm}{}^{-1}$. Notwithstanding, the computational
overhead of ab initio on-the-fly simulations demands for a further
reduction of the number of trajectories to be run.

This goal has been reached by means of the MC-SCIVR approach in which
rather than relying on a full Monte Carlo sampling of the phase space,
the SC time propagator is built using only a handful of tailored classical
trajectories. The trajectories are ideally selected according to the
Einstein-Brillouin-Keller (EBK) quantization rules

\begin{align}
\oint\limits _{H(\mathbf{Q}_{t}^{(n)},\mathbf{p}_{t}^{(n)})=E_{n}}p_{i}^{(n)}dq_{i}^{(n)}=\hbar\left(\nu_{i}+\frac{a_{i}}{2}+\frac{b_{i}}{4}\right),
\end{align}
where $\nu_{i}$ are positive integers, while $a_{i}$ and $b_{i}$
are Maslov indexes. Indeed the EBK quantization condition is exact
for the harmonic oscillator with Maslov indexes $a_{i}=1$ and $b_{i}=0$.
In this particular case

\begin{align}
\frac{1}{2}\left(p_{i}^{(n)}(t)\right)^{2}+\frac{1}{2}\omega_{i}^{2}\left(Q_{i}^{(n)}(t)\right)^{2}=\left(\frac{1}{2}+\nu_{i}\right)\hbar\omega_{i},\label{eq:EBK-quantization-t}
\end{align}
so that the classical trajectories have total energies (and energy
partition) corresponding to the harmonic oscillator spectral energies

\begin{align}
\mathcal{E}_{\pmb{\nu}}^{HO}=\sum_{i}\left(\frac{1}{2}+\nu_{i}\right)\hbar\omega_{i}.\label{eq:harm-spec}
\end{align}
The EBK quantization conditions are still exact when the PES anharmonicity
preserves the generalized periodicity of the motion (i.e. when each
mode performs a periodic motion). In fact, Eq.(\ref{eq:EBK-quantization-t})
and Eq.(\ref{eq:harm-spec}) become

\begin{align}
 & \sum_{i}\frac{1}{2}\left(p_{i}^{(n)}(t)\right)^{2}+V\left(\boldsymbol{Q}^{(n)}(t)\right)=\sum_{i}\left(\frac{1}{2}+\nu_{i}\right)~\hbar~\Xi_{i}(E)\nonumber \\
 & E_{n(\pmb{\nu})}=\sum_{i}~\left(\frac{1}{2}+\nu_{i}\right)~\hbar~\Xi_{i}(E_{n(\pmb{\nu})}),\label{eq:EBK-quantization-AN}
\end{align}
where $\Xi_{i}(E)$ are the classical frequencies of the generalized
periodic dynamics, which depend on the energy since the PES is not
harmonic.

The trajectories employed in SC dynamics require a short time evolution
(1 ps or less) without any preliminary equilibration. Therefore, the
harmonic EBK quantization conditions of Eq.(\ref{eq:EBK-quantization-t})
can be easily fulfilled at the initial step by setting

\begin{flalign}
Q_{i}^{n}(0) & =0\nonumber \\
p_{i}^{n}(0) & =\sqrt{(2\nu_{i}+1)\hbar\omega_{i}}.\label{eq:MC-initial-conds}
\end{flalign}
These conditions still realize a good approximation to the proper
EBK quantization condition of Eq.(\ref{eq:EBK-quantization-AN}) if
the energy dependence of the anharmonic frequencies is moderate enough,
i.e. if $\Xi_{i}(E_{n(\pmb{\nu})})\simeq\Xi_{i}(\mathcal{E}_{\pmb{\nu}}^{h})$.
The MC-SCIVR power spectrum is eventually computed as

\begin{flalign}
\tilde{I}_{\pmb{\nu},\textbf{\ensuremath{\chi}}}(E)\! & \propto\!Re\left[\int_{0}^{\tau}\!\!dt\braket{\chi|\hat{\mathcal{P}}_{n(\pmb{\nu})}^{MC}(t)|\chi}e^{\frac{i}{\hbar}Et}\right]=\nonumber \\
 & =\dfrac{1}{\tau}\left|\int_{0}^{\tau}dt\braket{\chi|\mathbf{Q}_{t}^{n},\mathbf{p}_{t}^{n}}e^{i[S_{t}(\mathbf{Q}_{0}^{n},\mathbf{p}_{0}^{n})+\phi_{t}(\mathbf{Q}_{0}^{n},\mathbf{p}_{0}^{n})+Et]/\hbar}\right|^{2}.\label{eq:MC_power_spectrum}
\end{flalign}
In calculations where $\ket{\chi}=\ket{\bf K}$ the term $\braket{\chi|\mathbf{Q}_{t}^{n},\mathbf{p}_{t}^{n}}$
is analytical.\citep{Micciarelli_Ceotto_SCwavefunctions_2018} We
have indicated with $n(\pmb{\nu})$ the $n$-th Hamiltonian eigenvalue
corresponding to the $\pmb{\nu}$ vector of integers via Eq.(\ref{eq:MC-initial-conds}).
The label $n(\pmb{\nu})$ has been added explicitly to the MC propagator
to indicate that it gives a reliable approximation to the exact propagator
only in the region of the energy spectrum close to $E_{n}$, i.e.

\begin{align}
\hat{\mathcal{P}}_{n(\pmb{\nu})}^{MC}(t)=\sum_{m}W_{n,m}^{2}~e^{-\frac{i}{\hbar}E_{m}^{(\mathcal{P}_{n})}t}\ket{e_{m}^{(\mathcal{P}_{n})}}\bra{e_{m}^{(\mathcal{P}_{n})}}
\end{align}
with

\begin{align}
E_{n}^{(\mathcal{P}_{n})}\simeq E_{n}\label{eq:MC-eigenenergies}
\end{align}
and

\begin{align}
W_{n}^{2}\ket{e_{n}^{(\mathcal{P}_{n})}}\simeq\ket{e_{n}},\label{eq:MC-eigenstates}
\end{align}
where the constants $W_{n}^{2}=W_{n,n}^{2}$ account for both the
potential loss of amplitude due to the separable approximation and
the fact that, having used a single trajectory, the amplitudes of
the different states are not converged uniformly.

In practice, the initial EBK conditions are chosen according to the
state to be investigated, and each different trajectory is used to
build a different propagator specialized in the energy range around
the eigenenergy of the target state. The MC-SCIVR approach has been
numerically tested in several applications returning accurate eigenvalues
and eigenvectors.\citep{Ceotto_AspuruGuzik_Multiplecoherent_2009,Micciarelli_Ceotto_SCwavefunctions_2018}
The typical strategy consists in obtaining a first estimate of the
SC frequencies of the fundamental transitions by means of the ground
state propagator. Calculations are then refined by employing a different
and tailored MC-SCIVR propagator (i.e. a different trajectory) for
each state. Using this approach, hence, the full SC power spectrum
can be composed piece after piece as a collection of different single-trajectory
propagators. In particular, for a given state $\ket{\chi}$, the MC-SCIVR
power spectrum obtained after a run of classical molecular dynamics
of length $\tau$ can be written as

\begin{align}
\tilde{I}_{\textbf{\ensuremath{\chi}}}(E\sim E_{n})\simeq\frac{1}{W_{n}^{2}}\tilde{I}_{\pmb{\nu},\textbf{\ensuremath{\chi}}}(E\sim E_{n})\simeq~|\braket{\chi|e_{n}}|^{2}\mathcal{D}(E-E_{n};\Gamma_{\tau}),\label{eq:MC-survival-general}
\end{align}
where, as already anticipated, the broadening functions $\mathcal{D}$
have the shape of squared sync functions. The constants $W_{n}^{2}$
can be eventually derived enforcing the normalization of the eigenstates,
i.e.

\begin{align}
1=\braket{e_{n}|e_{n}}=\sum_{\mathbf{K}}|C_{n,\mathbf{K}}|^{2}.
\end{align}
In fact, by selecting in Eq.(\ref{eq:MC-survival-general}) $\ket{\chi}=\ket{\phi_{\mathbf{K}}}$,
$E=E_{n}$, and $\mathcal{D}(0;\Gamma_{\tau})=1$ (see Appendix A
for a justification for this choice)

\begin{align}
W_{n}^{2}~=~\sum_{\mathbf{K}}\tilde{I}_{\pmb{\nu},\phi_{\mathbf{K}}}(E=E_{n}),\label{eq:Norm-en-MC}
\end{align}
which means that the constant for the $n$-th state can be calculated
as the sum of the (non-negligible) intensities at energy $E_{n}$
of all power spectra obtained using the harmonic states of the basis
set as reference states. For systems with sizable densities of states,
many different states and transitions may contribute to power spectra
and absorption bands. To avoid calculating all needed normalization
constants, they can be considered to be the same for all vibrational
states in the confidence energy range of each propagator. This is
justified by some tests which show that only a very mild discrepancy
($\sim1\%$) in the value of normalization constants comes from a
change in the reference energy position within the same confidence
energy window. This approximation is instead no longer valid when
different energy ranges and/or different propagators are taken into
consideration.

\section{RESULTS AND DISCUSSION\label{sec:RESULTS-AND-DISCUSSION}}

\textbf{H}\textsubscript{\textbf{2}}\textbf{O molecule} The first
test application we propose concerns the non-rotating water molecule
in vacuum. MC SCIVR was already applied to this system in our previous
study\citep{Micciarelli_Ceotto_SCwavefunctions_2018} and results
for both eigenenergies and eigenstates were in excellent agreement
with reference  calculations performed using the Grid Time-Dependent
Schrödinger Equation (GTDSE) computational package\citep{ssf+09a}.
From that work we borrowed the same, pre-existing analytical H\textsubscript{2}O
PES\citep{Dressler_Thiel_WaterPES_1997}, and the same pre-existing
dipole surface.\citep{tennyson_diph2o}

The initial step of any SC approach consists in providing a harmonic
estimate of vibrational frequencies. To this end, the Hessian matrix
at the equilibrium geometry has been diagonalized to get the three
harmonic frequencies of vibration, which are related to the symmetric
stretch ( $\omega_{s}=3831~\text{cm}^{-1}$), the bending ($\omega_{b}=1650~\text{cm}^{-1}$),
and the asymmetric stretch ($\omega_{a}=3941~\text{cm}^{-1}$) motions.
Consistently with the MC-SCIVR methodology presented above, to investigate
the 5 lowest-lying vibrational states we selected the appropriate
harmonic EBK initial conditions and then generated five classical
trajectories to build five MC-SCIVR propagators. The trajectories
were associated to the following triplets of harmonic quantum numbers
(in increasing order of energy): $\pmb{\nu}_{(n=1,\ldots,5)}=\left\{ (0,0,0);(0,1,0);(0,2,0);(1,0,0);(0,0,1)\right\} $.
Each trajectory was propagated for a total of 1.2 ps with Hessians
calculated at each step along the dynamics to evaluate the time evolution
of the Herman-Kluk prefactor (specifically its phase). We then applied
Eq.(\ref{eq:MC_power_spectrum}) to get 5 distinct MC-SCIVR power
spectra $\tilde{I}_{\pmb{\nu}_{n},\phi_{\pmb{\nu}}}(E)$. The final,
total power spectrum has been obtained as a direct sum of the 5 single
power spectra.

\begin{figure}
\begin{centering}
\includegraphics[width=150mm]{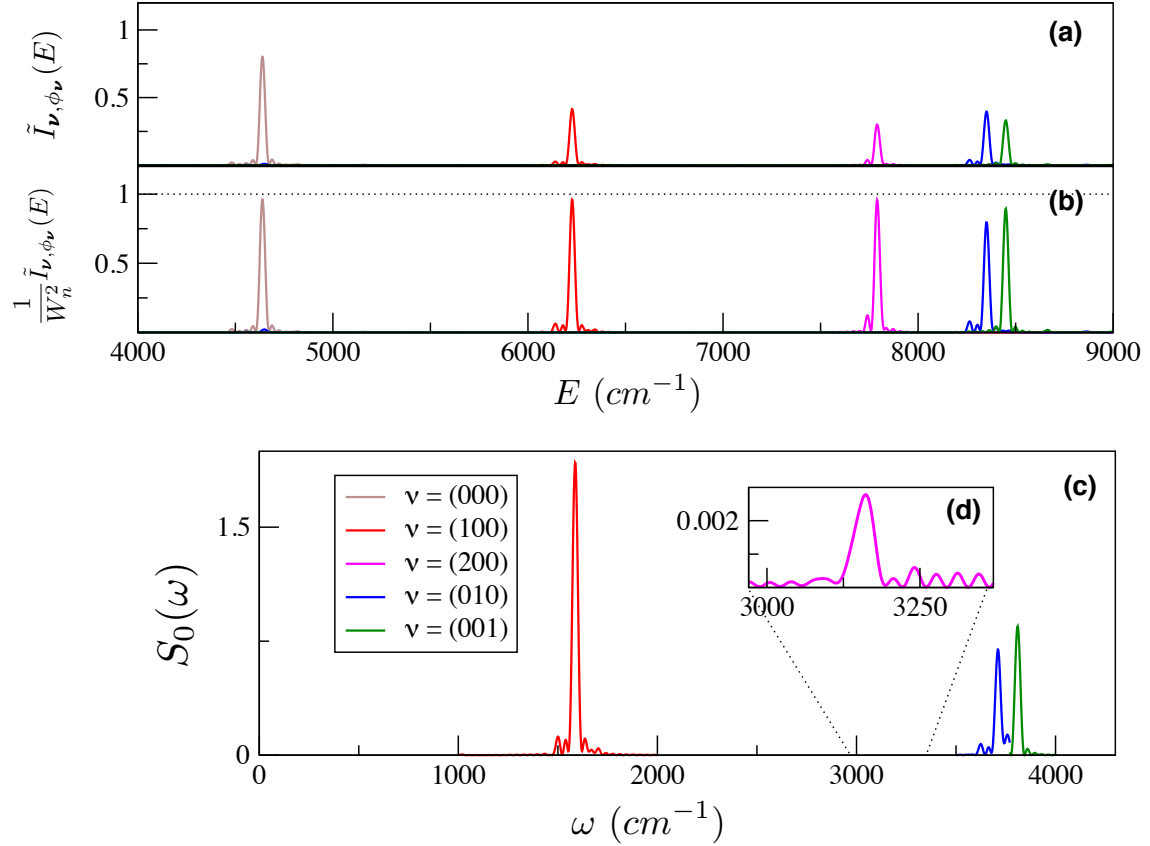}
\par\end{centering}
\centering{}\caption{Semiclassical power spectra for a non-rotating water molecule using
the MC approach based on single-trajectory propagators. The peaks
represent the ground state (brown), first bending excitation (red),
first bending overtone (magenta), and first symmetric (blue) and asymmetric
(green) excitations. In panel (a) power spectra from harmonic reference
states (i.e. $\pmb{\nu}=\mathbf{{K}}$) are reported. In panel (b)
the same spectra are shown upon normalization. Panel (c) illustrates
the semiclassical IR spectrum at 0K for unpolarized light. The inset,
panel (d), zooms in on the bending overtone.}
\label{fig:h2o_pes_spec}
\end{figure}

In panel (a) of Fig. \ref{fig:h2o_pes_spec} we report the total MC-SCIVR
power spectrum. As discussed above, the Hamiltonian eigenenergies
correspond to the positions of the different peaks. In panel (b) of
the same Figure we present the total power spectrum obtained from
the 5 normalized power spectra, i.e. $\frac{1}{W_{n}^{2}}\tilde{I}_{\pmb{\nu}_{n},\phi_{\pmb{\nu}}}(E)$.
The constants $W_{n}$ have been obtained from Eq.(\ref{eq:Norm-en-MC})
by employing a truncated harmonic basis set made of all possible harmonic
functions with quantum numbers not greater than 10, i.e. the total
number of harmonic states in the basis set was $11^{3}=1331$. The
different intensities of peaks between panel (a) and (b) in Fig. \ref{fig:h2o_pes_spec}
point out the importance of the normalization factors $W_{n}^{2}$
for this new SC approach very clearly. It is also worth noting that
the different peak intensities in panel (b) of Fig. \ref{fig:h2o_pes_spec}
are due to anharmonicity effects, which are related to the squared
projections $\left|\braket{\phi_{\pmb{\nu}_{n}}|e_{n}}\right|^{2}$.
However, as observed in our previous work,\citep{Micciarelli_Ceotto_SCwavefunctions_2018}
the main contribution to a generic anharmonic state $\ket{e_{n}}$
is given by its purely harmonic counterpart. Therefore all peak intensities
are indeed close to 1, since most of the character of each anharmonic
state is given by its harmonic associate. This is particularly true
for the ground and first bunch of excited states, while anharmonicity
increases in the excitation of bond stretches (corresponding to the
two peaks at the highest energies in Figure \ref{fig:h2o_pes_spec}).
For the latter, in fact, the normalized peak intensities become smaller.
The effect is larger for the symmetric stretch (blue curve) than for
the asymmetric one (green line). The reason is that a Fermi resonance
between the symmetric stretch $\ket{100}$ and the bending overtone
$\ket{020}$ is present.\\ In panel (c) of Fig. \ref{fig:h2o_pes_spec}
we show the fully anharmonic semiclassical IR spectrum of water under
the effect of unpolarized light ($S_{0}(E)=\sum_{\epsilon=x,y,z}S_{\epsilon,0}(E)$)
obtained using the ground state as a reference state. This is, hence,
the IR absorption spectrum at temperature $T=0$ K. The ground state
eigenfunction was expanded in terms of the harmonic states already
employed in our previous work,\citep{Micciarelli_Ceotto_SCwavefunctions_2018}
and the related coefficients employed in Eq.(\ref{eq:final-sc-ir-spec}).
We note that the intensity of the bending transition, located at $1587~\text{cm}^{-1}$,
is correctly almost twice as intense as the two stretching ones at
$3707~\text{cm}^{-1}$ and $3811~\text{cm}^{-1}$ respectively. Furthermore,
the bending overtone transition, estimated at $3162~\text{cm}^{-1}$,
is very weak but not exactly 0 because of the anharmonicity of the
PES.

\begin{figure}
\centering{}
\includegraphics[width=170mm]{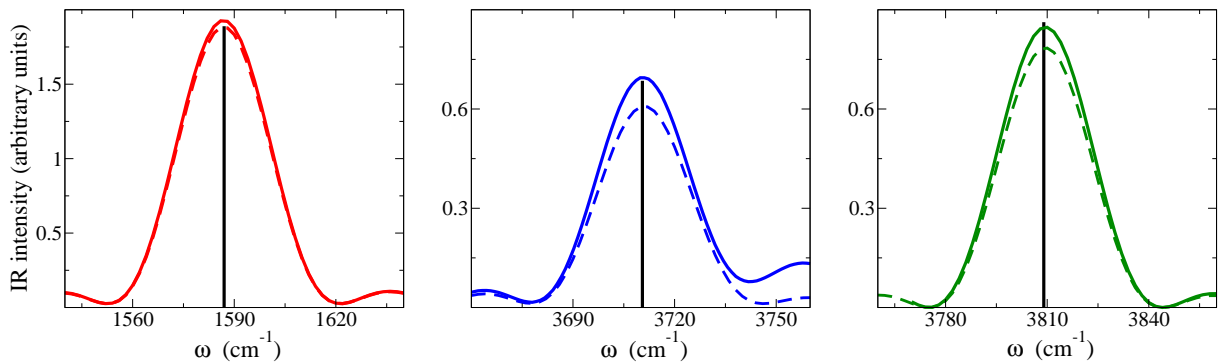} 
\caption{Detail of the three bright IR peaks of water reported in Fig. \ref{fig:h2o_pes_spec}.
The exact (vertical black line), SC semi-anharmonic (colored dashed
line), and SC fully anharmonic (colored solid line) intensity estimates
are reported in each plot.}
\label{fig:h2o_pes_spec_2}
\end{figure}

The exact anharmonic intensities of these absorption peaks have been
derived in our previous work by means of a DVR approach.\citep{Micciarelli_Ceotto_SCwavefunctions_2018}
They are reported as vertical black lines in Fig. \ref{fig:h2o_pes_spec_2}
for the bright vibrational fundamental transitions of water and compared
to the peaks of the SC IR spectrum. The agreement is excellent and
all three absorption intensities are perfectly reproduced within a
negligible error due to the dipole linearization and/or the semiclassical
approximation. Fig. \ref{fig:h2o_pes_spec_2} also points out the
enhanced accuracy of fully anharmonic IR spectra with respect to the
semi-anharmonic ones reported in dashed lines.

These results validate the proposed approach and show that it is equivalent
to the direct state-to-state calculation of the oscillator strengths
(i.e. the square moduli of transition dipoles). However, the decrease
in computational overhead is evident already at this low dimensionality.
In fact, the IR spectrum reported above asks only for the dipole derivative
with respect to nuclear displacements at the equilibrium geometry
in addition to what is needed to get a SC power spectrum. Any subsequent
dipole evaluation is not required.

\textbf{Glycine} We move to the 10-atom glycine molecule in its neutral
form in gas phase. Being the smallest among all aminoacids, this molecule
has both a great biological relevance and a manageable size, so several
theoretical methods have been applied to calculate its vibrational
spectrum beyond the harmonic approximation.\citep{Biczysko_Barone_abinitioIRgly_2012,Bludsky_Hobza_Anharmonicgly_2000,Senent_Fernandez_AbinitioIRgly_2005,Brauer_Gerber_Biologicalmolecules_2004,Stepanian_Adamowicz_IRandTheoreticalgly_1998,Barone_Puzzarini_ElusiveGly_2013}
In a recent work,\citep{Gabas_Ceotto_Glycine_2017} MC-SCIVR power
spectra have been calculated for the four main conformers of glycine
using on-the-fly ab initio molecular dynamics. Semiclassical energies
are in very good agreement with other theoretical calculations as
well as experimental data.\citep{Stepanian_Adamowicz_IRandTheoreticalgly_1998}
Here we extend the previous study by evaluating intensities and absorption
bands for the high-energy fundamental transitions, i.e. the CH\textsubscript{2}
and NH\textsubscript{2} stretches and the OH vibration, of the global
minimum conformer (Conf I). This spectroscopic region is of great
interest and the key target of investigation in bigger aggregates
because it is influenced by the hydrogen bonding responsible for the
structural stability of the complexes.\citep{Gabas_Ceotto_SupramolecularGlycines_2018}

\begin{figure}
\begin{centering}
\includegraphics[scale=0.3]{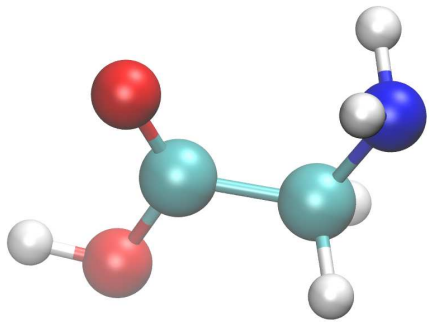}
\par\end{centering}
\caption{Ball-and-stick representation of the global minimum energy structure
of glycine in vacuum.}
\label{Fig:Gly-Ball-Stick}
\end{figure}

We performed ab initio on-the-fly molecular dynamics runs at DFT-B3LYP
level of theory with aug-cc-pVDZ basis set using the NWChem\citep{Valiev_DeJong_NWChem_2010}
suite of codes. The structure obtained for the global minimum conformer
is reported in ball-and-stick representation in Fig. \ref{Fig:Gly-Ball-Stick}.
The molecular dipole derivatives, which are necessary for calculating
the harmonic transition dipoles, the set of $24$ normal mode coordinates
and the harmonic spectrum have been computed at this molecular geometry
with the same level of theory and basis set. Starting from the double
harmonic approximation thus obtained, as anticipated, we focused on
the anharmonic corrections to the highest energetic fundamentals in
the energy range between $3000\ \text{cm}^{-1}$ and $4000\ \text{cm}^{-1}$,
for which harmonic estimates are known to be inaccurate.

\begin{figure}
\begin{centering}
\includegraphics[scale=0.6]{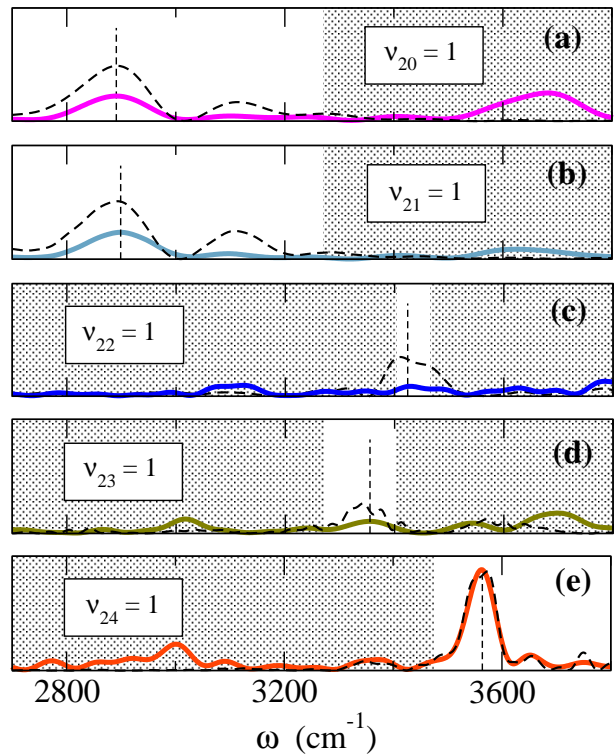}
\par\end{centering}
\caption{The semi-anharmonic IR absorption spectra of glycine in vacuum $S_{\phi_{0}}(\omega)$
(continuous color lines) are reported together with survival amplitude
power spectra $\tilde{I}_{\phi_{{\bf K}}}(\omega)$ (black dash lines)
with harmonic reference states $\phi_{\boldsymbol{K}}$ corresponding
to the EBK single trajectory used for the semiclassical propagator
(i.e. $\pmb{\nu}=\mathbf{{K}}$). Results have been obtained using
the propagator associated to the first harmonic excitations of modes
20-24 (panels from (a) to (e)). Dashed vertical lines are located
at the frequencies of the maxima of the absorption bands corresponding
to the fundamental transitions. For each propagator, the energy confidence
window is derived by the comparison of power and IR spectra and is highlighted
by reporting it in a region with white background.}
\label{Fig:GLY-specs-harm-e0}
\end{figure}

In Fig. \ref{Fig:GLY-specs-harm-e0} we report the five unpolarized
light semi-anharmonic IR spectra $S_{\phi_{\boldsymbol{0}}}(\omega)$
obtained using five different EBK trajectories. For each trajectory,
the initial conditions were determined by means of the harmonic EBK
quantization rule with vectors \textbf{$\boldsymbol{\nu}$ }obtained
giving one quantum of excitation to modes 20-24 as indicated in the
legend of the different figure panels. All trajectories have been
evolved for 5000 time steps $(dt=10$ $a.u.)$ giving a total evolution
time of $\sim1.2$ $ps$. It is worth mentioning that, according to
Eq.(\ref{eq:spec-Km-sc-finale}), for glycine the total number of
power spectra to be computed in order to get the semi-anharmonic IR
spectrum $S_{\phi_{0}}(\omega)$ would add up to $N_{v}(N_{v}+1)/2=300$
for each trajectory (i.e. for each SC propagator). However, for many
of the 24 single-excited harmonic reference states the power spectra
$\tilde{I}_{\phi_{\boldsymbol{0}^{(\alpha)}}}$ appearing in Eq.(\ref{eq:spec-Km-sc-finale})
give no contribution in the energy range of interest. Furthermore,
in these cases the $\Delta\tilde{I}_{\phi_{\boldsymbol{0}^{(\alpha)}},\phi_{\boldsymbol{0}^{(\beta)}}}$
terms vanish for all $\beta$, thus decreasing substantially the total
number of power spectra to be evaluated. In practice, the sole contributors
to the high energy bands investigated are the six modes from $19$
to $24$, decreasing the total number of power spectra calculated
to just 21.

Interestingly, energy shifts of the order of $100\ \text{cm}^{-1}$
and variations in band shapes are observed among different spectra,
but, consistently with the MC-SCIVR recipe, we consider each IR spectrum
reliable only within a given confidence energy range. These ranges
are selected in a way that they contain the absorption band located
where vibrational states have a significant component on the harmonic
state corresponding to the EBK trajectory adopted. The energy windows
are revealed by considering the (principal) band of the survival probability
power spectrum $\tilde{I}_{\phi_{\boldsymbol{K}}}(\omega)$, with
$\boldsymbol{K}=\boldsymbol{\nu}$ (reported with black dashed lines
in each panel) and have been highlighted in Fig. \ref{Fig:GLY-specs-harm-e0}.
As it can be seen from panel (a) and (b) of Fig. \ref{Fig:GLY-specs-harm-e0},
the confidence regions relative to modes 20 and 21 (corresponding
to the symmetric and asymmetric CH\textsubscript{2} stretches) coincide.
Furthermore, the two IR spectra obtained are equivalent within method
accuracy, and hence any of the two is representative of the IR spectrum
in that specific energy region, i.e. it is the result of the sum of
contributions from both CH\textsubscript{2} absorption bands. A different
situation applies in the case of the two $\text{NH}_{2}$ stretches,
whose confidence regions and absorption bands are not equal. Specifically,
data shown in panel (c) and (d) are obtained from the EBK trajectories
for the symmetric and asymmetric stretch, respectively. From the survival
probability power spectra, two non-overlapping confidence windows
can be determined. In panel (c) a double peak structure is present
and we assign the symmetric stretch to the peak at lower energy. In
panel (d) a single band conceals both states and its peak is taken
as the reference for the absorption of the asymmetric stretch fundamental,
in agreement with the underlying EBK trajectory. In this way, though,
the intensity of this band is expected to be overestimated because
contributions from the symmetric stretch are also included. However,
it turns out that in this specific case they are small. Finally, a
well defined single band characterizes the OH stretch fundamental
in panel (e).

As discussed in the theory Section and already pointed out for water,
the relative intensities of the different bands in Fig. \ref{Fig:GLY-specs-harm-e0}
are not directly comparable to each other in absence of a preliminary
normalization. Each IR spectrum in Fig. \ref{Fig:GLY-specs-harm-e0}
has been globally normalized by applying Eq.(\ref{eq:Norm-en-MC})
at the energy corresponding to the maximum of the absorption band
in the confidence region (i.e. in correspondence of the vertical dashed
lines reported in Fig. \ref{Fig:GLY-specs-harm-e0}). The relative
intensities of absorption bands in confidence regions obtained with
different propagators are hence meaningful. In order to determine
the normalization constants via Eq.(\ref{eq:Norm-en-MC}), the sum
over $\mathbf{K}$ has been performed over a truncated set of harmonic
basis set elements. The truncation strategy adopted consisted in considering
all possible 24-dimensional direct products of 1-dimensional harmonic
eigenstates up to the total quantum number $k_{max}=6$ with a maximum
of 2 modes simultaneously excited. The truncated basis set obtained
in this way contains $10081$ harmonic states. Not surprisingly, looking
at the amplitude of the expansion coefficients associated to each
peak, in all cases the largest coefficient belongs to the harmonic
excited state corresponding to the absorption band. However, dozens
of other relevant contributions are also present and many of them
are due to (dark) harmonic states in resonance with the principal
one. So, while within the harmonic approximation at T=0 K only fundamental
transitions give non-zero contributions to the IR spectrum, multiple
Fermi resonances mixing the bright harmonic states with many others
imply that the number of non-dark states becomes much larger in the
real anharmonic case. The absorption bands obtained in Fig. \ref{Fig:GLY-specs-harm-e0},
hence, are made of multiple energetically close vibrational transitions.
This is different from the simpler picture of absorption bands as
made of a single broadening function around a central bright transition.
Any state-by-state approach would fail to describe this phenomenology,
unless all oscillator strengths relative to transitions from a given
reference state to all states under the absorption band are taken
into account. Apart from the computational overhead required, this
procedure would be numerically unfeasible.

\begin{figure}
\begin{centering}
\includegraphics[scale=0.55]{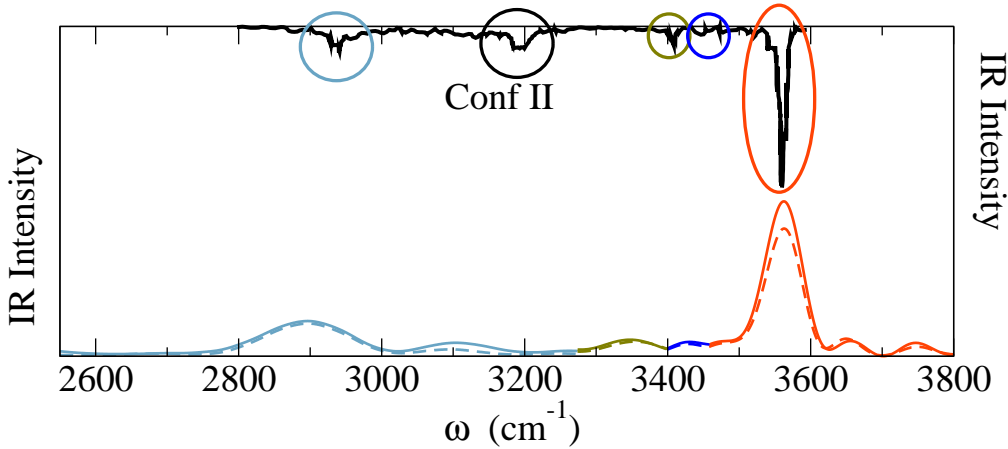}
\par\end{centering}
\caption{Comparison between experiment (black line, top) and the fully anharmonic
zero temperature IR spectrum of glycine in vacuum (continuous lines,
bottom). The semi-anharmonic spectrum (dashed lines, bottom) is also
reported. The fully and semi-anharmonic spectra have been obtained
by means of single trajectory propagators, each one valid in its confidence
energy range. This is indicated by the different colors chosen in
agreement with Fig. \ref{Fig:GLY-specs-harm-e0}.}
\label{Fig:Gly-specs-final}
\end{figure}

In Fig. \ref{Fig:Gly-specs-final} we report our estimate for the
zero temperature fully anharmonic IR spectrum $S_{0}(\omega)$. In
the same figure, the semi-anharmonic spectrum $S_{\phi_{0}}(\omega)$
is also presented in dashed lines. The different colors (the same
adopted in Fig. \ref{Fig:GLY-specs-harm-e0}) identify the single
trajectory propagator used in each confidence energy range and, for
all bands reported, the relative intensities are meaningful. The spectrum
$S_{0}(\omega)$ has been obtained by sorting the elements of the
harmonic basis set in order of relevance (i.e. absolute value) of
the corresponding ground state expansion coefficients and then applying
Eq.(\ref{eq:final-sc-ir-spec}) with the sum over harmonic states
performed over the first $50$ basis set elements. With this cutoff,
all coefficients with absolute value bigger than $\sim0.03$ in the
ground state expansion (see Fig. S1 in Supporting Information) are
considered. However, in order to keep under control the number of
power spectra to evaluate, we have limited the sum over normal mode
contributions to the same set of coordinates used for the calculation
of $S_{\phi_{0}}(\omega)$. The total number of power spectra taken
into account in this way, in fact, already adds up to $\sim2.6\times10^{6}$.
As shown in Fig. S2 of the Supporting Information, inclusion of more
and more harmonic contributions does have an observable effect on
the intensity of all bands with convergence obtained only after the
inclusion of the $40^{th}$ basis set element. The observed increment
in absorption intensities, however, is proportionally almost equivalent
for all bands. Hence, after a global re-normalization, the overall
effects of taking into account the anharmonicity of the vibrational
ground state (in terms of change in relative band intensities) become
minor.

It is worth noting that for absorption bands composed of more than
one transition line, the maximum may not formally coincide with the
maximum of a given survival amplitude power spectrum. A not negligible
change in the estimate of the position of the maximum of absorption
bands is indeed observed for the semi-anharmonic spectra of glycine
in panels (a)-(d) of Fig. \ref{Fig:GLY-specs-harm-e0}. Furthermore,
an additional even if less significant shift of most band maxima is
observed $(\lesssim5\ \text{cm}^{-1})$ when ground state anharmonicity
is also taken into account (see Fig. S2 in Supporting Information).
In Fig. \ref{Fig:Gly-specs-final} we report also the experimental
spectrum recorded in Argon matrix at low temperature.\citep{Stepanian_Adamowicz_IRandTheoreticalgly_1998}
A satisfactory overall agreement between our theoretical prediction
and the experimental spectrum is observed. The key difference is in
the region between $3150$ and $3250$ $\text{cm}^{-1}$, where the
experimental signature of another conformer, not included in our theoretical
investigation, stands out. The most intense absorption band is centered
at $3560$ $\text{cm}^{-1}$ and associated to the $\text{OH}$ stretch.
The second most intense experimental band is located in the energy
range between $2900$ and $3000$ $\text{cm}^{-1}$, with an overall
absorption intensity $\sim6$ times smaller than the OH one. This
band features a barely distinguishable bimodal shape and is associated
to the two (symmetric and asymmetric) $\text{CH}_{2}$ stretches.
The bimodality of this band is not resolved in our semiclassical spectrum
in which the broadening functions resulting from the dynamics are
indeed too large to recover this effect. Notably, however, a secondary
much less intense band located between $3050$ and $3150$ $\text{cm}^{-1}$
appears in the computed spectrum. In fact, as revealed by the power
spectra reported in Fig. \ref{Fig:GLY-specs-harm-e0} (black dashed
curves), in this spectral energy range vibrational states with relevant
components on the harmonic excited states of the two $\text{CH}_{2}$
stretches are present, most probably originated by the presence of
Fermi resonances coupling them with (dark) harmonic excited states.
The intensity of this secondary band, as expected, is much reduced
compared to the principal one in the semi-anharmonic spectrum, but
its oscillator strength increases upon inclusion of ground state anharmonicities
making it non negligible. However, the signal is not enough intense
to be assigned in the experimental spectrum and to be discernible
from noise or signals coming from other conformers. Finally, another
absorption band (assigned to the $\text{NH}_{2}$ stretches) in the
experimental spectrum spans the energy range between $3370$ and $3480$
$\text{cm}^{-1}.$ Its intensity, even if an order of magnitude smaller
than that of the OH stretch, makes it clearly distinguishable from
the experimental baseline and well matched in position and intensity
by the semiclassical prediction.

\begin{table}[H]
\begin{centering}
\begin{tabular}{ccccccccccccccc}
\cmidrule{1-1} \cmidrule{3-6} 
{\footnotesize{}Assignment} &  & \multicolumn{4}{c}{{\footnotesize{}MC-TA-SCIVR}} &  &  &  &  &  &  &  &  & \tabularnewline
{\footnotesize{}(Modes)} &  & \multicolumn{2}{c}{{\small{}Harmonic}} & \multicolumn{2}{c}{{\small{}Anharmonic}} &  &  &  &  &  &  &  &  & \tabularnewline
 &  & \multicolumn{4}{c}{{\footnotesize{}(B3LYP/aug-cc-pVDZ)}} &  &  &  &  &  &  &  &  & \tabularnewline
\cmidrule{3-6} 
 &  & {\footnotesize{}\phantom{aaaa}}{\small{}$\omega$}{\footnotesize{}\phantom{aaaa}} & {\footnotesize{}\phantom{aaaa}}{\small{}$I$}{\footnotesize{}\phantom{aaaa}} & {\footnotesize{}\phantom{aaaa}}{\small{}$\omega$}{\footnotesize{}\phantom{aaaa}} & {\footnotesize{}\phantom{aaaa}}{\small{}$I$}{\footnotesize{}\phantom{aaaa}} &  &  &  &  &  &  &  &  & \tabularnewline
\cmidrule{1-1} \cmidrule{3-6} \cmidrule{14-15} 
{\small{}$\text{CH}_{2}$ (20;21)} &  & {\scriptsize{}3051; 3089} & {\scriptsize{}0.08; 0.03} & {\scriptsize{}2900} & {\scriptsize{}0.07} &  &  &  &  &  &  &  & \multicolumn{2}{c}{{\footnotesize{}Expt.}}\tabularnewline
{\small{}$\text{NH}_{2}$ (22;23)} &  & {\scriptsize{}3495; 3568} & {\scriptsize{}0.01; 0.02} & {\scriptsize{}3345; 3430} & {\scriptsize{}0.03; 0.01} &  &  &  &  &  &  &  &  & \tabularnewline
\cmidrule{14-15} 
{\small{}OH (24)} &  & {\scriptsize{}3735} & {\scriptsize{}0.25} & {\scriptsize{}3563} & {\scriptsize{}0.25} &  &  &  &  &  &  &  & {\footnotesize{}\phantom{aa}}{\small{}$\omega$}{\footnotesize{}\phantom{aa}} & {\footnotesize{}\phantom{aa}}{\small{}$I$}{\footnotesize{}\phantom{aa}}\tabularnewline
\cmidrule{1-1} \cmidrule{3-6} \cmidrule{14-15} 
 &  & \multicolumn{4}{c}{{\footnotesize{}GVPT2}} &  &  &  &  &  &  &  & {\scriptsize{}$2935$} & {\scriptsize{}0.04}\tabularnewline
 &  & \multicolumn{2}{c}{{\small{}Harmonic}} & \multicolumn{2}{c}{{\small{}Anharmonic}} &  &  &  &  &  &  &  & {\scriptsize{}$3410;3450$} & {\scriptsize{}0.03;0.01}\tabularnewline
 &  & \multicolumn{4}{c}{{\footnotesize{}(B3LYP/N07D)}} &  &  &  &  &  &  &  & {\scriptsize{}$3560$} & {\scriptsize{}0.25}\tabularnewline
\cmidrule{3-6} \cmidrule{14-15} 
 &  & {\footnotesize{}\phantom{aaaa}}{\small{}$\omega$}{\footnotesize{}\phantom{aaaa}} & {\footnotesize{}\phantom{aaaa}}{\small{}$I$}{\footnotesize{}\phantom{aaaa}} & {\footnotesize{}\phantom{aaaa}}{\small{}$\omega$}{\footnotesize{}\phantom{aaaa}} & {\footnotesize{}\phantom{aaaa}}{\small{}$I$}{\footnotesize{}\phantom{aaaa}} &  &  &  &  &  &  &  &  & \tabularnewline
\cmidrule{1-1} \cmidrule{3-6} 
{\small{}$\text{CH}_{2}$ (20;21)} &  & {\scriptsize{}3044; 3079} & {\scriptsize{}0.08; 0.03} & {\scriptsize{}2938; 2929} & {\scriptsize{}0.09; 0.04} &  &  &  &  &  &  &  &  & \tabularnewline
{\small{}$\text{NH}_{2}$ (22;23)} &  & {\scriptsize{}3509; 3582} & {\scriptsize{}0.01; 0.03} & {\scriptsize{}3387; 3407} & {\scriptsize{}0.01; 0.02} &  &  &  &  &  &  &  &  & \tabularnewline
{\small{}OH (24)} &  & {\scriptsize{}3750} & {\scriptsize{}0.25} & {\scriptsize{}3568} & {\scriptsize{}0.25} &  &  &  &  &  &  &  &  & \tabularnewline
\cmidrule{1-1} \cmidrule{3-6} 
\end{tabular}
\par\end{centering}
\caption{Main features of the SC IR spectrum of glycine most stable conformer
in the high energy range are compared to experiments\citep{Stepanian_Adamowicz_IRandTheoreticalgly_1998}
as well as second order Generalized Vibrational Perturbation Theory
(GVPT2) calculations performed at the B3LYP/N07D level of theory.\citep{Biczysko_Barone_abinitioIRgly_2012}
Energy of maximum/maxima associated to each IR band are reported in
$\text{cm}^{-1}$ while intensities are scaled in order to level the
intensity of the $\text{OH}$ stretch band up to the experimental
one in all cases. Results from the double harmonic approximation are
also reported.}

\label{TAB:IR-GLY-comparisons}
\end{table}

Energies and intensities obtained from the semiclassical IR spectrum
are reported in Table \ref{TAB:IR-GLY-comparisons} and compared to
both experimental values and theoretical estimates provided by the
double harmonic approximation and by perturbative methods at a similar
level of electronic structure theory.\citep{Biczysko_Barone_abinitioIRgly_2012}
In order to facilitate the comparison, all intensities have been globally
re-normalized by leveling off the intensities assigned by the different
approaches to the intense $\text{OH}$ stretch band. This reference
value is set equal to the experimental value of 0.25. From Table \ref{TAB:IR-GLY-comparisons}
it is clear that a double harmonic approximation is not efficient
especially in estimating the frequencies of vibration, while a better
prediction is obtained for intensities. Results are neatly improved
moving to the SC and GVPT2 calculations. Both methods are in excellent
agreement with the experiment for all modes, taking alternatively
the lead as the most accurate approach. This demonstrates that the
SC approach works appropriately and accurately if compared to benchmark
calculations in both low (water) and high (glycine) dimensionality,
opening the way to the quantum simulation of IR spectra of systems
currently not achievable.

\section{Summary and Conclusions}

We have presented a new semiclassical approach to simulate anharmonic
IR spectra. The method is able to deal with medium-large molecular
systems with sizable densities of vibrational states. At the heart
of the strategy are MC-SCIVR power spectra, which are combined linearly
to give, in addition to frequencies of vibration, IR intensities and
band shapes.

Meaningful IR intensities are indeed obtained exploiting a proper
decomposition of the absorption spectrum in terms of survival amplitude
power spectra. As for band shapes the method relies on the fact that,
as the number of absorption lines under a given band becomes large
and in the limit in which the width of the band is much larger than
the width of the single lines, the shape of the band is independent
of the broadening function adopted. This allows to employ the convolution
function associated to the finite-time survival amplitude Fourier
transform as a broadening function in lieu of the usual Gaussian or
Lorentzian envelopes. Furthermore, this feature permits to avoid undesired
state-to-state oscillator strength calculations, whose number becomes
huge for large systems making their computation hardly feasible. In
fact, even though MC-SCIVR power spectra need a proper normalization
to allow comparisons between different energy regions, normalization
constants refer to the central peak of absorption bands and barely
vary within it. Thus, only one constant per absorption band has to
be computed, a figure that in large systems is much smaller than the
actual number of lines composing each band.

The approach is also based on the linearization of the molecular dipole,
but intensities of absorption spectra can still be calculated at the double harmonic,
semi-anharmonic, or fully anharmonic level. The difference depends
on the way the states contributing to the linearized transition dipoles
are treated. The accuracy of the method has been tested against exact
results for the water molecule with more than satisfactory outcome.
Furthermore, the study of the IR spectrum of glycine has provided
results in agreement with both experiments and previous VPT2 calculations
and demonstrated the effectiveness of the method in dealing with a
24-dimensional system. Investigation of glycine focused on the absorption
bands of high energy fundamentals, an energy region where anharmonicities
are relevant. The importance of anharmonicities at both electrical
and mechanical level is evident after a comparison to the widely employed
double harmonic approximation. In both cases explored, the double
harmonic approximation is off the mark, which raises serious questions
about the legitimacy of employing such a rough guess in a black box
fashion.

A still open issue is that, due to the short duration of the semiclassical
propagation, the SC absorption bands obtained for glycine are wider
than the low-temperature experimental ones. This effect could be less
relevant for spectra at higher temperature, where experimental bands
are expected to extend over a larger interval of energies. Furthermore,
the quality of the SC calculation could be improved by reducing the
impact of spurious rotations, which may contribute to the enlargement
of peaks and are due to the adoption of a normal mode reference frame.
Advances on this aspect are currently being undertaken.

We conclude by remarking that SC spectroscopy, through power spectra
calculations, had already well established two of its hallmark features.
They are the possibility of application to large dimensional molecular
systems, which is often precluded to other quantum approaches, and
the capability to reproduce and explain experimental findings where
other theoretical approaches (like scaled harmonic or classical ones)
fail. Now we have increased the appeal of semiclassical spectroscopy
by demonstrating that semiclassical IR spectra are also achievable
without any restrictions due to the density of states. This gives
SC approaches the potential to be reference methods for IR spectroscopy
of medium-large systems.

\section*{SUPPLEMENTARY MATERIAL}

See supplementary material for additional data about the expansion
coefficients of the ground state vibrational eigenfunction of glycine
and an additional plot about convergence of the full-anharmonic IR
spectrum of glycine as a function of the number of expansion coefficients
taken into account.
\begin{acknowledgments}
Authors acknowledge financial support from the European Research Council
(Grant Agreement No. (647107)\textemdash SEMICOMPLEX\textemdash ERC-
2014-CoG) under the European Union\textquoteright s Horizon 2020 research
and innovation programme. Additional cpu time was provided by CINECA
(Italian Supercomputing Center) under ISCRAB project ``QUASP''.
\end{acknowledgments}

\numberwithin{equation}{section}

\appendix

\section{}

From quantum linear response theory in its sum-over-state version,
the functional form of the IR spectrum of isotropic and homogeneous
molecular systems is

\begin{align}
S(\omega,T)=\sum_{n}\sum_{m\neq n}[P_{n}(T)-P_{m}(T)]~\Omega_{nm}~\frac{1}{3}~|\pmb{\mu}_{nm}|^{2}~\delta(\omega-\Omega_{nm}),\label{eq:abs-spec}
\end{align}
where $\Omega_{nm}=E_{m}-E_{n}$ is the difference between the vibrational
energies of a given transition, $\pmb{\mu}_{nm}=\braket{e_{n}|\hat{\pmb{\mu}}_{0N}|e_{m}}$
is the corresponding transition dipole, and $P_{l}=e^{-\frac{E_{l}}{k_{B}T}}/Z$
is the $l^{th}$ vibrational state population at a given temperature
T (with $Z=\sum_{n}e^{-\frac{E_{n}}{k_{B}T}}$ being the partition
function). In this work we are interested in the modeling of the IR
spectrum in the region of the high energy fundamentals $(\omega\gtrsim2500~cm^{-1})$
of a system at a given (reasonable) temperature $T$. Hence, $\Omega_{nm}\gg K_{B}T$
and it can be assumed that all the arrival state populations $P_{m}(T)$
are zero. Eq.(\ref{eq:abs-spec}) simplifies to

\begin{align}
S(\omega,T) & =\sum_{n}\sum_{m\neq n}P_{n}(T)~\Omega_{nm}~\frac{1}{3}~|\pmb{\mu}_{nm}|^{2}~\delta(\omega-\Omega_{nm})=\nonumber \\
 & =\sum_{n}P_{n}(T)S_{n}(\omega),\label{eq:abs-spec-high-E}
\end{align}
where

\begin{align}
S_{n}(\omega) & =\sum_{m\neq n}~\Omega_{nm}~\frac{1}{3}~|\pmb{\mu}_{nm}|^{2}~\delta(\omega-\Omega_{nm}).\label{eq:spec-n}
\end{align}

In experimental IR spectra each spectral line is broadened over a
finite range of energies. This is due to several known effects. For
instance thermal Doppler, collisional, and Stark effect broadening
all play an important role in a gas phase environment. These effects
can be accounted for phenomenologically by assuming that the transition
probability distributions of the system are broadened, so that in
Eq.(\ref{eq:spec-n}) the Dirac delta function can be substituted
by some bell shaped function $\mathcal{L}(\omega-\Omega_{nm};\Gamma)$
of finite amplitude $\Gamma$. Eq. (\ref{eq:spec-n}) in this case
becomes

\begin{align}
S_{n}(\omega)=\sum_{m\neq n}~\frac{1}{3}~|\pmb{\mu}_{nm}|^{2}~\omega~\mathcal{L}(\omega-E_{m}+E_{n};\Gamma).\label{eq:spec-n-bro-1}
\end{align}
It is straightforward to notice from a comparison between Eq. (\ref{eq:spec-n})
and Eq. (\ref{eq:spec-n-bro-1}) that the term $\Omega_{nm}$, characteristic
of a transition between two single and well defined states, has been
substituted by $\omega$ in the case of an absorption band. This is
necessary to account for the gain of a quantum $\hbar\omega$ of energy
by the system (and, correspondingly, for the loss of an equal amount
of energy by the electromagnetic field) in the neighborhood of $\Omega_{nm}$
with probability density given by $\frac{1}{3}~|\pmb{\mu}_{nm}|^{2}~\mathcal{L}(\omega-E_{m}+E_{n};\Gamma)$.
Depending on the most relevant phenomenological effect to take into
account, $\mathcal{L}$ is generally described by either a Gaussian
or a Lorentzian function. In this work $\mathcal{L}$ is substituted
by the dynamical convolution function $\mathcal{D}(\omega-E_{m}+E_{n};\Gamma_{\tau})$,
a legitimate procedure as explained in the Summary and Conclusions
Section of the paper. Therefore the calculated spectrum in presence
of a radiation polarized along the direction $\epsilon$, $S_{\epsilon,n}(\omega)$,
descends from Eq.(\ref{eq:spec-n-bro}).

It is common practice to use normalized broadening functions that
integrate to unity over the energy axis. This choice, in fact, has
the advantage to preserve the oscillator strength summation rules.
The most convenient choice for our purposes, instead, is to normalize
them by setting $\mathcal{D}(0,\Gamma_{\tau})=$1. This choice has
the advantage to allow a straightforward calculation of the normalization
constants of the eigenstates expanded on the harmonic basis via Eq.(\ref{eq:Norm-en-MC}).
These two normalization choices are in any case equivalent because
the resulting absorption spectra just differ in a multiplicative constant.

\numberwithin{equation}{section}

\section{}

We derive the decomposition of the fully anharmonic IR spectrum of
Eq.(\ref{eq:spec-n-bro}) in terms of power spectra, i.e. the result
reported in Eq.(\ref{eq:final-sc-ir-spec}). First, Eq.(\ref{eq:dip-with-sel-rules})
for transition dipoles has to be inserted into Eq.(\ref{eq:spec-n-bro}),
leading to

\begin{align}
S_{\epsilon,n}(\omega)=\omega\sum_{m\ne n}~\left|\sum_{\mathbf{K}}C_{n,\mathbf{K}}\sum_{\alpha}C_{m,\mathbf{K}^{(\alpha)}}\mu_{\mathbf{K}^{(\alpha)}}^{\epsilon}\right|^{2}\mathcal{D}(\omega-E_{m}+E_{n};\Gamma_{\tau}).\label{eq:spec-n-SC-1}
\end{align}
At this point the square modulus is expanded to allow inversion of
the order of the sums over $\alpha$ and over $m$. From an algebraic
point of view this procedure is not straightforward because of the
presence of multiple sums, so we perform it starting from the following
general expansion

\begin{align}
\left|\sum_{\mathbf{K}}C_{n,\mathbf{K}}A_{m,\mathbf{K}}\right|^{2}=\sum_{\mathbf{K}}\Bigg[C_{n,\mathbf{K}}^{2}A_{m,\mathbf{K}}^{2}+\sum_{\mathbf{K'}<\mathbf{K}}2C_{n,\mathbf{K}}C_{n,\mathbf{K'}}A_{m,\mathbf{K}}A_{m,\mathbf{K'}}\Bigg].
\end{align}
If
\begin{equation}
A_{m,\mathbf{K}}=\sum_{\alpha}C_{m,\mathbf{K}^{(\alpha)}}\mu_{\mathbf{K}^{(\alpha)}}^{\epsilon},
\end{equation}
 then 
\begin{equation}
A_{m,\mathbf{K}}A_{m,\mathbf{K'}}=\sum_{\alpha,\alpha^{\prime}}C_{m,\mathbf{K}^{(\alpha)}}C_{m,\mathbf{K'}^{(\alpha')}}\mu_{\mathbf{K}^{(\alpha)}}^{\epsilon}\mu_{{\mathbf{K'}^{(\alpha')}}}^{\epsilon},
\end{equation}
and upon substitution into Eq.(\ref{eq:spec-n-SC-1})

\begin{align}
S_{\epsilon,n}(\omega)= & \omega\sum_{\mathbf{K}}C_{n,\mathbf{K}}^{2}\Bigg\{\sum_{\alpha}\Bigg[\bigg(\sum_{m\neq n}C_{m,\mathbf{K}^{(\alpha)}}^{2}\mathcal{D}(\omega-E_{m}+E_{n})\bigg)(\mu_{\mathbf{K}^{(\alpha)}}^{\epsilon})^{2}+\nonumber \\
+ & \sum_{\beta<\alpha}\mu_{\mathbf{K}^{(\alpha)}}^{\epsilon}\mu_{\mathbf{K}^{(\beta)}}^{\epsilon}\bigg(\sum_{m\neq n}2C_{m,\mathbf{K}^{(\alpha)}}C_{m,\mathbf{K}^{(\beta)}}\mathcal{D}(\omega-E_{m}+E_{n})\bigg)\Bigg]+\nonumber \\
+ & \sum_{{\mathbf{K'}}<{\mathbf{K}}}\sum_{\alpha,\alpha^{\prime}}C_{n,\mathbf{K}}C_{n,\mathbf{K'}}\bigg(\sum_{m\neq n}2C_{m,\mathbf{K}^{(\alpha)}}C_{m,\mathbf{K'}^{(\alpha')}}\mathcal{D}(\omega-E_{m}+E_{n})\bigg)\mu_{\mathbf{K}^{(\alpha)}}^{\epsilon}\mu_{\mathbf{K}^{(\alpha')}}^{\epsilon}\Bigg\},\label{eq:spec-n-SC-2}
\end{align}

The final Eq.(\ref{eq:final-sc-ir-spec}) is obtained starting from
Eq.(\ref{eq:spec-n-SC-2}) by noting that: i) the terms in brackets
are the semi-anharmonic spectra $S_{\mathbf{\epsilon,\phi_{K}}}(E)$;
ii) in the last line, the terms in parentheses are also easily related
to survival amplitudes of the kind of Eq.(\ref{eq:surv-harm-states}).

\bibliographystyle{aipnum4-1}
\bibliography{Scarrafone2}

\end{document}